# The Road Map toward Room-Temperature Superconductivity: Manipulating Different Pairing Channels in Systems Composed of Multiple Electronic Components

**Annette Bussmann-Holder** [1,*], **Jürgen Köhler** [1], **Arndt Simon** [1], **Myung-Hwan Whangbo** [2], **Antonio Bianconi** [3,4,5] **and Andrea Perali** [6,3]

[1] Max-Planck-Institute for Solid State Research, Heisenbergstr. 1, D-70569 Stuttgart, Germany;
[2] Department of Chemistry, North Carolina State University, Raleigh, 27695-8204 NC, USA;
[3] Rome International Center for Materials Science Superstripes (RICMASS), Via dei Sabelli 119A, 00185 Rome, Italy
[4] Institute of Crystallography, Consiglio Nazionale delle Ricerche CNR, Via Salaria Km 29.300, 00015 Monterotondo, Italy
[5] National Research Nuclear University, MEPhI (Moscow Engineering Physics Institute), Kashirskoye sh. 31, Moscow 115409, Russia
[6] School of Pharmacy, Physics Unit, University of Camerino, 62032 Camerino, Italy

While it is known that the amplification of the superconducting critical temperature $T_C$ is possible in a system of multiple electronic components in comparison with a single component system, many different road maps for room temperature superconductivity have been proposed for a variety of multicomponent scenarios. Here we focus on the scenario where the first electronic component is assumed to have a vanishing Fermi velocity corresponding to a *case of the intermediate polaronic regime,* and the second electronic component is in the weak coupling regime with standard high Fermi velocity using a mean field theory for multiband superconductivity. This roadmap is motivated by compelling experimental evidence for one component in the proximity of a Lifshitz transition in cuprates, diborides, and iron based superconductors. By keeping a constant and small exchange interaction between the two electron fluids, we search for the optimum coupling strength in the electronic polaronic component which gives the largest amplification of the superconducting critical temperature in comparison with the case of a single electronic component.

**1.INTRODUCTION**

Room-temperature superconductivity has been, and is the dream of, scientists, since the discovery of superconductivity in 1911, not only due to its enormous potential in applications, but also for the experimental evidence of the possibility of emergence of macroscopic quantum coherence at room temperature. After the discovery of high-temperature superconductivity in cuprate perovskites [1] the record $T_C$ of 164 K was held by $HgBa_2Ca_{m-1}Cu_mO_{2m+2+\delta}$ with $m = 3$ and at high pressure [2]. In 2015 Eremets and his collaborators succeeded in observing superconductivity in sulfur hydride with a very high $T_C$ of 203 K at an ultrahigh pressure of 150 GPa [3] which has been confirmed experimentally by several groups [4–7] and it is the object of high theoretical interest [8–10].

The phenomenon of room-temperature superconductivity is frequently associated with strong coupling and real space pairing approaching a BEC regime, contrasted with weak coupling and momentum space pairing of conventional low $T_C$ BCS superconductors. For the BCS superconductor the distance between paired electrons can be of the order of several thousands of Angstroms, much larger than the average inter-particle distance whereas, for high-temperature superconductors, this distance is of the order of some unit cells with formation of local molecular-like pairs. From the viewpoint of theory in the strong coupling limit the electron-lattice system enters the polaronic regime with low charge density and low Fermi energy, where phonon frequencies reach an ultralow energy. However, in this regime superconductivity competes with the proximity to a structural lattice instability and localization of charge carriers [11]. While conventional BCS superconductors are stable, high-temperature superconductors have an inherent tendency for structural instabilities, as discussed by Sleight [12].

In agreement with the prediction [1], polarons have been detected in cuprates [13–15], using X-ray absorption spectroscopy [16,17] and confirmed by theories [18,19]. However, experiments show two unexpected features.

First, the polaronic carriers are in a coupling regime intermediate [13–15] between large polarons in the weak coupling, expected to condense in the BCS fashion, and the small polarons in the strong coupling regime, localized in a single site and expected to form bosonic bipolarons which should condense into the BEC regime. The intermediate sized polarons in the in between lying coupling regime in cuprates extend over a distorted nanoscale domain of about 8 unit cells, [13–15] and are supposed to condense in the BCS-BEC crossover regime [20], which occurs upon increasing the pairing coupling strength in the pathway from momentum space pairing (BCS regime) to real space pairing (BEC regime).

Second, the cuprates are characterized by two electronic components [21–23] however while the most popular scenario of high $T_C$ superconductivity in a two component system has been the boson-fermion model made of BEC and BCS condensates [13], it was found that the electronic phase of cuprates consists of a first electronic component in the intermediate coupling regime condensing the BCS-BEC crossover which coexists with a second quasi-free electron component in the weak coupling regime [14,15,19,21] which is expected to condense in the BCS limit.

The key role of configuration interaction between a first pairing channel in the BCS-BEC sector and a second pairing channel in the BCS regime has been proposed by Bianconi–Perali–Valletta (BPV) for cuprates [24–26], and a road map for material design of high-temperature superconductors with multiple electronic components has been suggested in [27–30].

In the BPV proposed scenario [24–30] the maximum amplification of the superconducting critical temperature is predicted to occur where two conditions should be realized in a scenario called superstripes: first, the materials are composed of superlattices of quantum wells or quantum wires and, second, the chemical potential is tuned at a Lifshitz transition or near a band edge where a "shape resonance" emerges between pairing in the first BEC-BCS condensate and the second BCS condensate. In the first component the ratio of the its Fermi energy $E_{F1}$ to the characteristic pairing energy $\omega_0$, is about $E_{F1}/\omega_0 = 1.5$ and the pairing is in the regime of the BCS-BEC crossover, while the pairing in the second electronic component is in the BCS regime, where the Fermi energy $E_{F2}$ is a large energy scale, $E_{F2}/\omega_0 \gg 1$. The amplification of the critical temperature $T_C$ in comparison with the predictions of the standard single band BCS approximation is optimized for weak coupling strengths, in the $\lambda_{22}$ pairing channel and weak exchange interband pairing interaction $\lambda_{12}$, where $\lambda_{ii}, \lambda_{ij}$ correspond to effective coupling constants. In the very strong coupling scenario in these two channels the amplification factor becomes smaller. Therefore, the road map for room temperature superconductivity based on the BPV approach predicts an optimal case where a first electronic system being at particular values of the polaronic regime in the $\lambda_{11}$ channel, intermediately between weak and strong coupling. It is interesting to remark that the vertex corrections near the band edge proportional to $\lambda_{11} \approx \text{sqrt}(\omega_0/E_{F1})$ are not expected to be large for a coupling constant $0.1 < \lambda_{11} < 0.35$ and the ratio $\omega_0/E_{F1}$ close to 1.

In cuprates it has been confirmed that multiple electronic components are involved in the pairing from penetration depth [31,32], ARPES [33,34], gap-to-$T_C$ ratio [35], the doping-dependent isotope effect [36–39] measurements, and the interband exchange interaction has been put forward to play a crucial role in enhancing $T_C$, as well as in explaining these key experiments [36–41]. Therefore, the recipe to realize stable room temperature superconductivity has been suggested for complex systems composed of two different electronic components where instabilities are avoided [27–29,36–38] through the interactions between them, the polaronic component and the free particles, contributing to a single superconducting phase. The scenario where the unconventional superconductivity emerges from the coexistence of a flat band and a steep band was proposed to occur in the case of superconductivity in $CaC_6$ [42,43] in contrast to elemental calcium.

In 2001 the predicted BPV scenario was shown to emerge clearly in superconducting $MgB_2$, where in a superlattice of atomic boron layers intercalated by magnesium layers a wide band of boron π electrons coexists with boron σ electrons with a small Fermi energy, since the chemical potential is near the σ band edge in the energy scale of its energy fluctuations due to zero point motion [44–49]. A very large amplification of the critical temperature appears as a function of the shift of the chemical potential below the σ band edge. It has been explained by Bussmann-Holder et al. [50] using the multiband approach of Suhl et al. [51], by Ummarino et al. [52] using the multi-band Eliashberg theory, and by the Innocenti et al. [53] using the BPV theory.

The case of magnesium diboride provided a clear confirmation that, *first*, the high $T_C$ materials should be formed in a nanoscale superlattice of atomic layers and/or second that multiple Fermi surface spots should coexist and give different superconducting condensates for each electronic component. Finally, the fine tuning of the chemical potential should be achieved in different ways by pressure, by chemical doping, or by misfit strain of charge injection (or by a combination of these) in order to cross an electronic topological transition (ETT) called a Lifshitz transition [54,55], which is a type of quantum criticality [56]. In fact, the tuning of the chemical potential near a band edge, called a Lifshitz transition for the appearance of a new Fermi surface spot, leads to resonant phenomena between different pairing channels. In particular the shape resonances between superconducting gaps are Fano resonances or Feshbach resonances between the superconducting first condensate in the BCS-BEC crossover regime and second condensates in the BCS regime.

In 2008 the discovery of iron based superconductors was rapidly recognized to be another realization of a superstriped material where the Fermi level is close to a Lifshitz transition [53,57–60]. In fact they present an alternative implementation of a heterostructure at the atomic limit [27–29] made of a superlattice of metallic quantum wells (FeAs or FeSe). It was soon confirmed that these materials are multiband superconductors since some of the five Fe 3d-orbitals cross the chemical potentials and create multiple Fermi surfaces. While the majority of the scientific community accepted the theoretical paradigm s+/- of Fermi surface nesting between equivalent Fermi surfaces of the similar area, other scientists proposed the shape resonance scenario since 2009–2011 [53,57-60]. In this scenario the large Fermi surfaces coexist with an incipient small Fermi surface since the chemical potential is fine-tuned near its Lifshitz transition for the appearing of a new Fermi surface. The superconductivity critical temperature variation by tuning the chemical potential was predicted to occur in the proximity of the emergence of a corrugated cylindrical quasi 2D Fermi surface of one of the five Fe 3d orbitals [53,57-60].

It was necessary to improve the energy resolution of ARPES experiments down to a few meV and to tune the photon energy to get the top of the incipient band dispersing in the c-axis direction. Finally, since 2011 several high-resolution ARPES experiments have experimentally confirmed that the maximum critical temperature occurs near a Lifshitz transition [61–67] and now, in 2017, the s+/- paradigm has been falsified and the Lifshitz transition paradigm [68] is widely accepted by the majority of scientists in the field.

For pressurized sulfur hydride, while the first proposed paradigm in 2015 was "superconductivity in a single effective band" [8], the two components' superconductivity scenario has been proposed [9,10], namely, that at least two bands are involved in the pairing mechanism, with one band being rather polaronic in character whereas the other one is itinerant [69–72]. This corresponds closely to the steep band/flat band model, and the shape resonance paradigm where a first BCS-BEC crossover-like condensate and a second BCS pairing condensate coexist in the same superconducting phase. The isotope coefficient shows rather large variations with pressure [69–72] as a function of doping analogous to cuprates, with a sizable peak where the chemical potential crosses the Lifshitz transition. Therefore, we invoke this scenario also for sulfur hydrides since its band structure provides evidence that a flat band is located at the Fermi energy in coexistence with steep bands reminiscent of MgB2 and cuprates.

In this work we propose a road map based on our knowledge of different realizations of the shape resonance scenario. We consider the first band to be in the intermediate coupling regime represented by a first electronic component, the intermediated polaron liquid, with vanishing Fermi velocity and a second electronic component, the quasi-free particles, with large Fermi velocity and in the weak coupling BCS regime. We fix the exchange-term controlling the pair exchange interaction, represented by the interband pairing, at a low value so that the mixing between the two condensates is small and the two condensates are distinguishable with quite different gap/$T_C$ ratio as in the experiments [35,53].

## 2. RESULTS AND DISCUSSION

The Hamiltonian for our model is given by:

$$H = H_1 + H_2 + H_3 \tag{1}$$

$$H_1 = \sum_{k,\sigma} [\varepsilon_1(k) a^+_{k,\sigma} a_{k,\sigma} + \varepsilon_2(k) b^+_{k,\sigma} b_{k,\sigma}] \tag{2}$$

$$H_2 = -\frac{1}{V} \sum_{k \neq g} [V_1 a^+_{k\uparrow} a^+_{-k\downarrow} a_{-g\downarrow} a_{g\uparrow} + V_2 b^+_{k\uparrow} b^+_{-k\downarrow} b_{-g\downarrow} b_{g\uparrow}] \tag{3}$$

$$H_{12} = -\frac{1}{V} \sum_{k \neq g} [V_{12}(a^+_{k\uparrow} a^+_{-k\downarrow} b_{-g\downarrow} b_{g\uparrow} + b^+_{k\uparrow} b^+_{-k\downarrow} a_{-g\downarrow} a_{g\uparrow})] \tag{4}$$

Here, $V$ is the volume of the system and the terms in $H_i$ ($i = 1 - 3$) are momentum $k$, $g$ dependent. Electron creation and annihilation operators are denoted $a^+, a$ in band 1 and $b^+, b$ in band 2. The effective intraband pairing potentials are given by $V_1, V_2$, whereas $V_{12} = V_{21}$ stems from interband pair scattering which is assumed to be the same for scattering between channels 1 and 2 and 2 and 1. The band energies $\epsilon_i(k)$ reflect the localized flat band and the itinerant one and are correspondingly approximated by: $\varepsilon_1 = const. = B$; $\varepsilon_2(k) = k^2/2m$. More complex and realistic band dispersions have been considered, e.g., for cuprates [32,36], since, here, the emphasis is on a flat band (polaronic)/steep band (itinerant) scenario, we use the simplest dispersion form to mimic this.

After performing a Bogoliubov transformation of the Hamiltonian, the finite temperature mean-field gap equations are explicitly obtained $\Delta^*_1 = \sum_k V_1 \langle a^+_{k\uparrow} a^+_{-k\downarrow} \rangle$ and analogous for $\Delta^*_2$, together with $A^*_k = \sum_k V_{12} \langle b^*_{k\uparrow} b^*_{-k\downarrow} \rangle$; $B^*_k = \sum_k V_{12} \langle a^*_{k\uparrow} a^*_{-k\downarrow} \rangle$, through which the order parameters are coupled. This yields the self-consistent set of coupled gap equations:

$$\Delta_1 = \frac{V_1}{V} \sum_k \frac{\Delta_1}{\Omega_1(k)} tanh\left[\frac{\Omega_1(k)}{2kT}\right] + \frac{V_{12}}{V} \sum_k \frac{\Delta_2}{\Omega_2(k)} tanh\left[\frac{\Omega_2(k)}{2kT}\right] \tag{5}$$

$$\Delta_2 = \frac{V_2}{V} \sum_k \frac{\Delta_2}{\Omega_2(k)} tanh\left[\frac{\Omega_2(k)}{2kT}\right] + \frac{V_{12}}{V} \sum_k \frac{\Delta_1}{\Omega_1(k)} tanh\left[\frac{\Omega_1(k)}{2kT}\right] \tag{6}$$

with $\Omega_i(k) = \sqrt{\varepsilon_i(k)^2 + \Delta_i^2}$ being the quasiparticle excitation energies for each component in the superconducting phase. These equations have to be solved simultaneously and self-consistently for each temperature $T$ in order to derive the temperature dependence of the coupled gaps. Since a variation of the chemical potential does not enter explicitly in our model system, the effect of doping is simulated by varying the inter/intraband pairing potentials. The critical temperature $T_C$ is given by the condition $\Delta_1, \Delta_2$ to yield the linearized gap equations:

A. Bussman-Holder et al. arXiv:1704.00276

$$\Delta_1 = \frac{V_1}{V}\sum_k \frac{\Delta_1}{\varepsilon_1(k)}\tanh\left[\frac{\varepsilon_1(k)}{2kT_c}\right] + \frac{V_{12}}{V}\sum_k \frac{\Delta_2}{\varepsilon_2(k)}\tanh\left[\frac{\varepsilon_2(k)}{2kT_c}\right] \quad (7)$$

$$\Delta_2 = \frac{V_2}{V}\sum_k \frac{\Delta_2}{\varepsilon_2(k)}\tanh\left[\frac{\varepsilon_2(k)}{2kT_c}\right] + \frac{V_{12}}{V}\sum_k \frac{\Delta_1}{\varepsilon_1(k)}\tanh\left[\frac{\varepsilon_1(k)}{2kT_c}\right] \quad (8)$$

Upon replacing the summations by integrals and introducing the density of states at the Fermi level $N_i(0)$, the dimensionless coupling constants are defined by $\lambda_{ii} = N_i(0)V_i$, $\lambda_{12} = \sqrt{N_1(0)N_2(0)}V_{12}$. With this definition the above equations become:

$$1 = \lambda_{11}\int_0^{\hbar\omega_1}\frac{1}{\varepsilon_1(k)}\tanh\left[\frac{\varepsilon_1(k)}{2kT_c}\right]d\varepsilon_1 + \lambda_{12}\sqrt{\frac{N_2}{N_1}}\int_0^{\hbar\omega_{1,2}}\frac{1}{\varepsilon_2(k)}\tanh\left[\frac{\varepsilon_2(k)}{2kT_c}\right]d\varepsilon_2 \quad (9)$$

$$1 = \lambda_{22}\int_0^{\hbar\omega_2}\frac{1}{\varepsilon_2(k)}\tanh\left[\frac{\varepsilon_2(k)}{2kT_c}\right]d\varepsilon_2 + \lambda_{12}\sqrt{\frac{N_1}{N_2}}\int_0^{\hbar\omega_{1,2}}\frac{1}{\varepsilon_1(k)}\tanh\left[\frac{\varepsilon_1(k)}{2kT_c}\right]d\varepsilon_1 \quad (10)$$

where $\omega_1, \omega_2, \omega_{1,2}$ refer to the cutoff frequencies related to band 1,2 whereas $\omega_{1,2}$ is either $\omega_1$ or $\omega_2$, depending on the band considered in the respective integration. By considering, explicitly, the band energies given above, attributing a large, but finite, density of states also for the flat band, and defining $\sqrt{N_1/N_2} = C$, Equations (9,10), they can be reformulated like:

$$1 = \lambda_{11}\int_0^{\hbar\omega_1}\frac{1}{B}\tanh\left[\frac{B}{2kT_c}\right]d\varepsilon_1 + \lambda_{12}\frac{1}{C}\int_0^{k_{1,2}}\frac{1}{k^2/2m}\tanh\left[\frac{k^2/2m}{2kT_c}\right]dk \quad (11)$$

$$1 = \lambda_{22}\int_0^{k_2}\frac{1}{k^2/2m}\tanh\left[\frac{k^2/2m}{2kT_c}\right]dk + \lambda_{21}C\int_0^{\hbar\omega_{1,2}}\frac{1}{B}\tanh\left[\frac{B}{2kT_c}\right]d\varepsilon_1 \quad (12)$$

with the parameter $B$ being calculated self-consistently to obey the same zero gap relations. The coupled integrals can be solved analytically for certain limits only, namely $x \gg 1$ where the $\tanh(x)$ approaches 1, or in the opposite limit $x \ll 1$ when it becomes $x$. All intermediate cases, as discussed below, have to be solved numerically.

The logarithmic BCS expression is used to approximate the integrals involving the second parabolic band, and with $\lambda_{12} = \lambda_{21}$ the integrals can be solved like:

$$1 = \lambda_{11}\int_0^{\hbar\omega_1}\frac{1}{B}\tanh\left[\frac{B}{2kT_c}\right]d\varepsilon_1 + \lambda_{12}\frac{1}{C}\int_0^{\hbar\omega_{1,2}}\frac{1}{\varepsilon_2(k)}\tanh\left[\frac{\varepsilon_2(k)}{2kT_c}\right]d\varepsilon_2 \quad (13)$$

$$1 = \lambda_{22}\int_0^{\hbar\omega_2}\frac{1}{\varepsilon_2(k)}\tanh\left[\frac{\varepsilon_2(k)}{2kT_c}\right]d\varepsilon_2 + \lambda_{12}C\int_0^{\hbar\omega_{1,2}}\frac{1}{B}\tanh\left[\frac{B}{2kT_c}\right]d\varepsilon_1 \quad (14)$$

$$1 = \frac{\lambda_{11}\hbar\omega_1}{B}\tanh\left[\frac{B}{2kT_c}\right] + \lambda_{12}\frac{1}{C}\ln\left[\frac{1.13\hbar\omega_{1,2}}{kT_c}\right] \quad (15)$$

$$1 = \lambda_{22}\ln\left[\frac{1.13\hbar\omega_2}{kT_c}\right] + \frac{\lambda_{12}C\hbar\omega_{1,2}}{B}\tanh\left[\frac{B}{2kT_c}\right] \quad (16)$$

The two approximate solutions are (i) $x \ll 1$: $\tanh(x) = x$; (ii) $x \gg 1$: $\tanh(x) = 1$.

$$(i)\ 1 = \lambda_{11}\frac{\hbar\omega_1}{2kT_c} + \lambda_{12}\frac{1}{C}\ln\left[\frac{1.13\hbar\omega_{1,2}}{kT_c}\right] = \lambda_{12}\,C\,\frac{\hbar\omega_{1,2}}{2kT_c} + \lambda_{22}\ln\left[\frac{1.13\hbar\omega_2}{kT_c}\right] \quad (17)$$

yielding an implicit relation for $T_C$.

$$(ii)\ 1 = \frac{\lambda_{11}\hbar\omega_1}{B} + \lambda_{12}\frac{1}{C}\ln\left[\frac{1.13\hbar\omega_{1,2}}{kT_c}\right] = \lambda_{12}C\frac{\hbar\omega_{1,2}}{B} + \lambda_{22}\ln\left[\frac{1.13\hbar\omega_2}{kT_c}\right] \quad (18)$$

which can be solved explicitly:

$$kT_c = [1.13\hbar\omega_2\,exp\left(\frac{-\lambda_{22}}{\lambda_{12}\frac{1}{C}-\lambda_{22}}\right) + 1.13\hbar\omega_{1,2}\,exp\left(\frac{-\lambda_{12}}{\lambda_{22}-\lambda_{12}C}\right)]exp\left(\frac{\lambda_{12}\frac{1}{C}\hbar\omega_{12}-\lambda_{11}\hbar\omega_1}{B(\lambda_{22}-\lambda_{12}C)}\right) \quad (19)$$

This expression differs from the one as discussed in the general two-band approach introduced by Suhl, Matthias, and Walker (SMW) [51] where the phonon cutoff energies are identical for the different pairing channels, and the band dispersion remains undefined. In the SMW approach the isotope effect is modified only through the exponential:

$$kT_c = 1.13\hbar\omega\, exp\left(-\frac{1}{\lambda}\right)$$

$$1/\lambda = \frac{1}{2}\left[\lambda_{11} + \lambda_{22} \pm \sqrt{(\lambda_{11} - \lambda_{22})^2 + 4\lambda_{12}^2/[\lambda_{11}\lambda_{22} - \lambda_{12}^2]}\right] \quad (20)$$

In the following the coupled gap equations are solved numerically using the assumption that a strong coupling and a weak coupling band coexist. In addition $C = 1$ because, for simplicity, we incorporate the effects of the different strengths of the pairing interaction in the coupling matrix. In this case one superconducting gap (strong coupling band) is substantially larger than the second one (weak coupling band). The flat/band steep band scenario admits two possibilities, namely, that the flat band is realized either in the strong coupling band or in the weak coupling band. For both cases the enhancement of $T_C$ caused by the interband interaction is calculated, as well as the temperature dependence of the related gaps and the isotope effects on $T_C$ which helps in understanding the experimental results and leads to conclusions concerning the mechanism realized in sulfur hydride.

Special cases of the above scenarios are studied in the following, where $\omega_1 = \omega_2 = 150\ meV$, namely, the combination: $\lambda_{22} = 0.1$, $\lambda_{12} = 0.03$, $\lambda_{22} = 0.2$, $\lambda_{12} = 0.03$, and $\lambda_{22} = 0.2$, $\lambda_{12} = 0.1$. In all three cases the only variable is now $\lambda_1$ related to the polaronic flat band which is varied from 0.05 to 0.6, i.e., from very weak to moderately strong coupling. The other parameters as given above are in the weak coupling limit, thus, allowing to use a BCS type scheme to evaluate the characteristic properties of our system.

First, $T_C$ is calculated as a function of $\lambda_{11}$ for the three cases given above (Figure 1a). An amazing and generic dependence of $T_C$ on $\lambda_{11}$ is clearly seen since $T_C$ remains negligibly small up to values of $\lambda_{11} = 0.2$ with the exception of the increased interband coupling $\lambda_{12} = 0.1$ which also induces for small values of $\lambda_{11}$ considerable increases in $T_C$. This does not hold for $\lambda_{22}$ which—when doubled—does not affect much the $T_C$ versus $\lambda_{11}$ dependence suggesting that its role is almost negligible. For all values $\lambda_{11} > 0.2$ a rapid increase in $T_C$ takes place to reach values of more than 300 K for $\lambda_{11} = 0.6$.

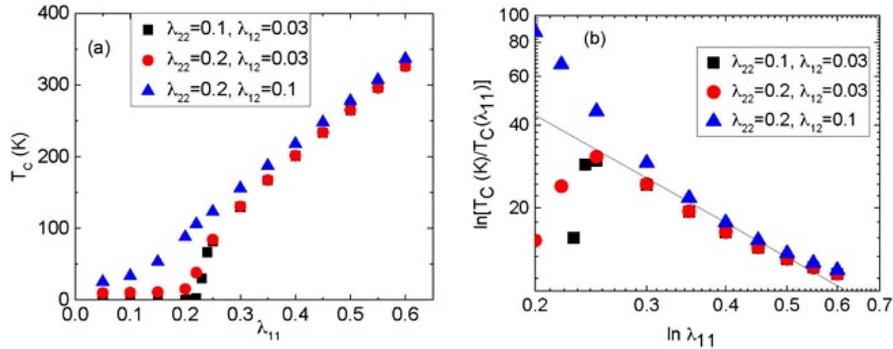

**Figure 1.** (**a**) $T_C$ as a function of the intraband coupling $\lambda_{11}$ which is related to the strong coupling flat polaronic band. The black squares refer to the parameters $\lambda_{22} = 0.1$, $\lambda_{12} = 0.03$, red circles: $\lambda_{22} = 0.2$, $\lambda_{12} = 0.03$, and blue triangles: $\lambda_{22} = 0.2$, $\lambda_{12} = 0.1$. (**b**) Double logarithmic plot of the normalized $T_C$ as a function of the interband coupling.

This observation demonstrates that a leading role in $T_C$ enhancements relates to the intraband polaron coupling. In order to evidence the increases in $T_C$ in better detail, the values shown in Figure 1a have been normalized to the single band $T_C$ values when only the polaronic band is considered (Figure 1b). While, for small $\lambda_{11}$, $T_C$ is very small and, correspondingly, the normalized value is large, a strong depression of $T_C$ with increasing $\lambda_{11}$ takes place highlighting its crucial role. All three cases follow an almost unique dependence for $\lambda_{11} \geq 0.25$.

The corresponding gaps and the dependence of them on $T_C$ and $\lambda_{11}$ are shown for the same parameter combinations in Figure 2a,b. Apparently, the larger polaron related superconducting gap increases in a unique manner as a function of $T_C$, whereas the second gap strongly depends on the interband interaction which induces strong increases in it with increasing $\lambda_{12}$, emphasizing its importance for the itinerant band. The BCS ratio $2\Delta_{1,2}/kT_C$ is shown in Figure 3a–c. As expected, the leading gap to $T_C$ ratio is enhanced as compared to the BCS value as long as $\lambda_{11} > 0.25$ whereas, for values $\lambda_{11} < 0.25$, the leading gap is reversed. This feature is generic and independent of the parameter choice. Their average remains, however, always smaller as compared to the BCS value, rather analogous to Al-doped MgB$_2$ [50,53,63].

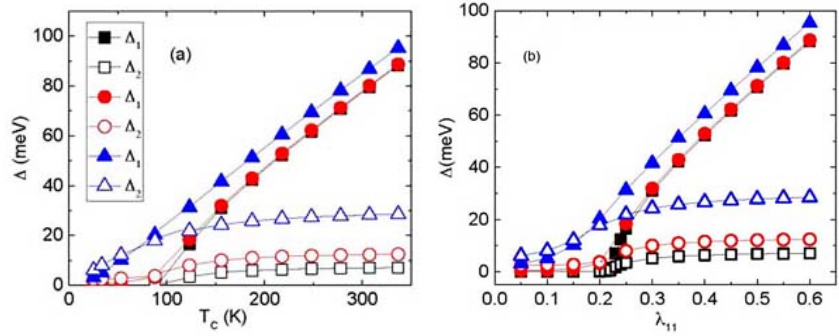

**Figure 2.** (**a**) The superconducting gaps $\Delta_i$ ($i$ = 1,2) as a function of the corresponding $T_C$. (**b**) The superconducting gaps as a function of the intraband coupling $\lambda_{11}$. (Symbols and colors refer to the same parameters as in Figures 1 and 2a.)

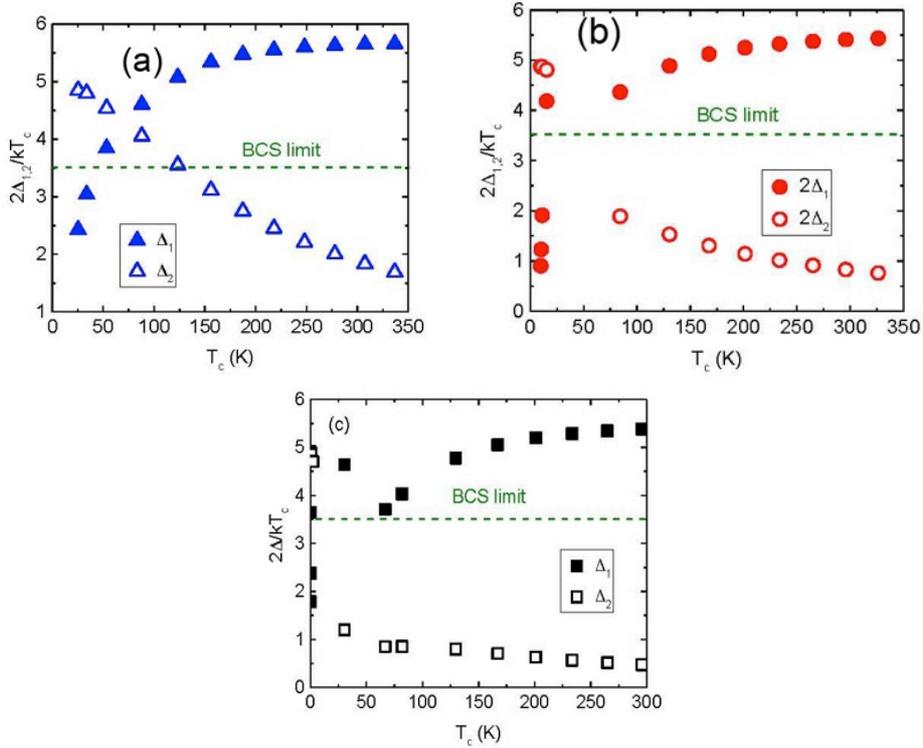

**Figure 3.** $2\Delta_{1,2}/kT_C$ as a function of $T_C$ for the three parameters sets discussed in the text. (**a**) $\lambda_{22}$ = 0.2, $\lambda_{12} = 0.1$. (**b**) $\lambda_{22} = 0.2$, $\lambda_{12} = 0.03$. (**c**) $\lambda_{22} = 0.1$, $\lambda_{12} = 0.03$.

The isotope effect resulting from the above model is shown in Figure 4a,b. Consistent with Figures 1 and 2, $\lambda_{11} = 0.2$ marks a crossover from almost BCS values to substantially reduced values smaller than 0.1.

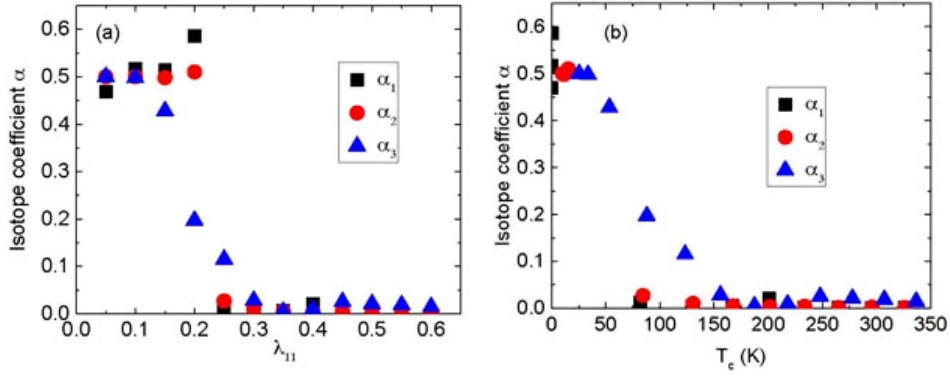

**Figure 4.** The isotope exponent $\alpha_{1,2,3}$ (1, 2, 3 refer to the three cases of coupling constant combinations discussed above) as a function of $\lambda_{11}$ panel (**a**) and of $T_C$ panel (**b**)

The crossover is characterized by the fact that the leading polaronic gap $\Delta_1$ adopts large values, whereas $\Delta_2$ shows only minor increases with increasing $T_C$, respectively, to almost saturate for $\lambda_{11} > 0.4$. This can be more clearly seen by plotting the gap ratio $\Delta_2/\Delta_1$ versus $\lambda_{11}$ as is done in Figure 5. In spite of the fact that $\Delta_2$ remains small even at large values of $T_C$, it is essential for enhancing $T_C$ since the interband coupling is a basic ingredient in the model.

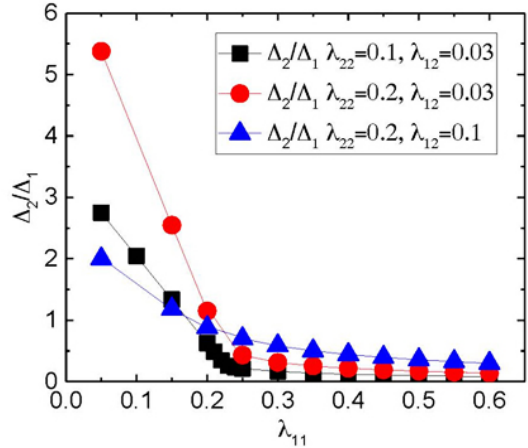

**Figure 5.** The gap ratio $\Delta_2/\Delta_1$ as a function of $\lambda_{11}$.

Again, a generic dependence for the gap ratios is observed for $\lambda_{11} > 0.25$ which marks a boundary value between different regimes. Below this value the three cases are rather different, where small interband couplings enhance the dominance of $\Delta_2$ as compared to $\Delta_1$, whereas an increased interband interaction removes it.

## 3. CONCLUSIONS

To conclude, high-temperature superconductivity in sulfur hydride and possible room-temperature superconductivity cannot be achieved within conventional single-band/single-gap theories in the sense of Eliashberg, BCS or strong coupling theories since all these approaches limit $T_C$ to values smaller than 30 K due to the inherent inverse relation between phonon energy and the electron-phonon coupling constant. In addition, isotope effects are not pressure dependent as long as the structure remains unchanged. even though the value of $\alpha$ can be smaller than the BCS value. The scenario outlined in this work shows that a two-band/two-gap theory can easily lead to values of $T_C$ > 200 K using rather moderate inter- and intraband coupling constants. By extending the SMW model to a combination of a flat band, i.e., dispersion-less polaronic type, and a steep itinerant one, moderate coupling is realized in combination with weak coupling. Such an approach resembles the BEC-BCS crossover discussed broadly in the literature for ultracold fermionic atoms and for the high-$T_C$ superconductors [73,74], however, with the distinction that not a single crossover is considered, but the coexistence of different pairing regimes in the two different bands. The strong coupling flat band can be related to polaron and bipolaron physics, whereas the itinerant electrons delocalize the localized ones through pair-exchange interband interactions. The isotope effect arising from this model can—when combined with experimental data—distinguish what scenario is realized in the respective compound. For sulfur hydride apparently a flat polaronic band combined with a steep itinerant band accounts for the observed experimental data. We have suggested that not only sulfur hydride, but also cuprates, pnictides, and MgB$_2$ are multiband systems where the coexistence of bosonic-like composite pairs (small Cooper pairs) in terms of bipolarons and fermions coupled through interband interactions are the important ingredients to increase $T_C$ to the observed large values. The special case of the combination of strong coupling polaronic and weak coupling itinerant case has been investigated in deeper detail by concentrating on the role of the intraband coupling related to the bosonic band. In contrast to the well-known SMW model, new physics emerge from this approach, namely, a considerable enhancement of $T_C$ caused by the intraband coupling only. Additinoally, the gap to $T_C$ ratios are substantially modified as compared to the standard SMW model. The isotope effect stemming from this combination decreases abruptly when the intraband couplings become pronouncedly different, indicating a crossover from BCS to coexistence of BCS and BCS-BEC crossover physics [75,76]. From the above results, certain directions in the search for new superconductors can be deduced, namely that layered compounds with distinctly different properties of the constituting layers are well suited to meet the above criteria. Further heterostructures at the atomic limit, so called superstripes materials like cuprates, diborides, and iron based superconductors, are ideal candidates to observe high-temperature superconductivity.

Finally we would like to note that these materials show an inherent inhomogeneity as it has been found experimentally [77–81], and theoretically it is expected near a Lifshitz transition in strongly interacting electron fluids with strong lattice fluctuations [82–86], which give complex geometries for the interplay of the multiple electronic components in high-temperature superconductivity with a key role of percolating superconducting pathways to establish macroscopic quantum coherence in complex systems [87,88].

**References**

1. Bednorz, J.G.; Müller, K.A. Perovskite-type oxides the new approach to high-Tc superconductivity. *Rev. Mod. Phys.* **1988**, *60*, 585–600, doi:10.1103/revmodphys.60.585.
2. Gao, L.; Xue, Y.Y.; Chen, F.; Xiong, Q.; Meng, R.L.; Ramirez, D.; Chu, C.W.; Eggert, J.H.; Mao, H.K. Superconductivity up to 164 K in HgBa$_2$Ca$_{m-1}$Cu$_m$O$_{2m+2+\delta}$ ($m$ = 1, 2, and 3) under quasi-hydrostatic pressures. *Phys. Rev. B* **1994**, *50*, 4260–4263, doi:10.1103/physrevb.50.4260.
3. Drozdov, A.P.; Eremets, M.I.; Troyan, I.A.; Ksenofontov, V.; Shylin, S.I. Conventional superconductivity at 203 Kelvin at high pressures in the sulfur hydride system. *Nature* **2015**, *525*, 73–76, doi:10.1038/nature14964.
4. Einaga, M.; Sakata, M.; Ishikawa, T.; Shimizu, K.; Eremets, M.I.; Drozdov, A.P.; Troyan, I.A.; Hirao, N.; Ohishi, Y. Crystal structure of the superconducting phase of sulfur hydride. *Nat. Phys.* **2016**, *12*, 835–838, doi:10.1038/nphys3760.
5. Goncharov, A.F.; Lobanov, S.S.; Kruglov, I.; Zhao, X.M.; Chen, X.J.; Oganov, A.R.; Konôpková, Z.; Prakapenka, V.B. Hydrogen sulfide at high pressure: Change in stoichiometry. *Phys. Rev. B* **2016**, *93*, 174105, doi:10.1103/physrevb.93.174105.
6. Huang, X.; Wang, X.; Duan, D.; Sundqvist, B.; Li, X.; Huang, Y.; Li, F.; Zhou, Q.; Liu, B.; Cui, T. Direct Meissner effect observation of superconductivity in compressed H$_2$S. *arXiv* **2016**, arXiv:1610.02630.

7. Goncharov, A.F.; Lobanov, S.S.; Prakapenka, V.B.; Greenberg, E. Stable high-pressure phases in the H-S system determined by chemically reacting hydrogen and sulfur. *arXiv* **2017**, arXiv:1702.02522.
8. Gor'kov, L.P.; Kresin, V.Z. Pressure and high-$T_C$ superconductivity in sulfur hydrides. *Sci. Rep.* **2016**, *6*, 25608.
9. Gordon, E.E.; Xu, K.; Xiang, H.; Bussmann-Holder, A.; Kremer, R.K.; Simon, A.; Köhler, J.; Whangbo, M.-H. Structure and composition of the 200 K-Superconducting phase of $H_2S$ at ultrahigh pressure: The perovskite ($SH^-$)($H_3S^+$). *Angew. Chem. Int. Ed.* **2016**, *55*, 3682–3684, doi:10.1002/anie.201511347.
10. Bianconi, A.; Jarlborg, T. Superconductivity above the lowest Earth temperature in pressurized sulfur hydride. *Europhys. Lett.* **2015**, *112*, 37001, doi:10.1209/0295-5075/112/37001.
11. Brosha, E.L.; Davies, P.K.; Garzon, F.H.; Raistrick, I.D. Metastability of Superconducting Compounds in the Y-Ba-Cu-O System. *Science* **1993**, *260*, 196–198, doi:10.1126/science.260.5105.196.
12. Sleight, A.W. Room temperature superconductors. *Acc. Chem. Res.* **1995**, *28*, 103–108. doi:10.1021/ar00051a003.
13. Müller, K.A.; Bussmann-Holder, A. (Eds.) *Superconductivity in Complex System*; Volume 114 of the Series Structure and Bonding; Springer: Berlin, Germany, 2005.
14. Bianconi, A.; Missori, M.; Oyanagi, H.; Yamaguchi, H.; Nishiara, Y.; Della Longa, S. The measurement of the polaron size in the metallic phase of cuprate superconductors. *Europhys. Lett.* **1995**, *31*, 411–415.
15. Bianconi, A.; Saini, N.L.; Lanzara, A.; Missori, M.; Rossetti, T.; Oyanagi, H.; Yamaguchi, H.; Oka, K.; Ito, T. Determination of the local lattice distortions in the $CuO_2$ plane of $La_{1.85}Sr_{0.15}CuO_4$. *Phys. Rev. Lett.* **1996**, *76*, 3412–3415, doi:10.1103/physrevlett.76.3412.
16. Bianconi, A.; Bachrach, R.Z. Al surface relaxation using surface extended X-ray-absorption fine structure. *Phys. Rev. Lett.* **1979**, *42*, 104–108, doi:10.1103/physrevlett.42.104.
17. Della Longa, S.; Soldatov, A.; Pompa, M.; Bianconi, A. Atomic and electronic structure probed by X-ray absorption spectroscopy: Full multiple scattering analysis with the G4XANES package. *Comput. Mater. Sci.* **1995**, *4*, 199–210, doi:10.1016/0927-0256(95)00027-n.
18. Bersuker, G.I.; Goodenough, J.B. Large low-symmetry polarons of the high-$T_C$ copper oxides: Formation, mobility and ordering. *Phys. C Supercond. Appl.* **1997**, *274*, 267–285, doi:10.1016/s0921-4534(96)00636-3.
19. Kusmartsev, F.V.; Di Castro, D.; Bianconi, G.; Bianconi, A. Transformation of strings into an inhomogeneous phase of stripes and itinerant carriers. *Phys. Lett. A* **2000**, *275*, 118–123, doi:10.1016/s0375-9601(00)00555-7.
20. Guidini, A.; Flammia, L.; Milošević, M.; Perali, A. BCS-BEC crossover in quantum confined superconductors. *Supercond. Nov. Magn.* **2016**, *29*, 711–715, doi:10.1007/s10948-015-3308-y.
21. Bianconi, A. On the Fermi liquid coupled with a generalized Wigner polaronic CDW giving high $T_C$ superconductivity. *Solid State Commun.* **1994**, *91*, 1–5, doi:10.1016/0038-1098(94)90831-1.
22. Muller, K.A. Possible coexistence of *s*- and *d*-wave condensates in copper oxide superconductors. *Nature* **1995**, *377*, 133–135. doi:10.1038/377133a0.
23. Müller, K.A.; Zhao, G.-M.; Conder, K.; Keller, H. The ratio of small polarons to free carriers in derived from susceptibility measurements. *J. Phys. Condens. Matter* **1998**, *10*, L291–L296, doi:10.1088/0953-8984/10/18/001.
24. Perali, A.; Bianconi, A.; Lanzara, A.; Saini, N.L. The gap amplification at a shape resonance in a superlattice of quantum stripes: A mechanism for high $T_C$. *Solid State Commun.* **1996**, *100*, 181–186, doi:10.1016/0038-1098(96)00373-0.
25. Valletta, A.; Bianconi, A.; Perali, A.; Saini, N.L. Electronic and superconducting properties of a superlattice of quantum stripes at the atomic limit. *Z. Phys. B Condens. Matter* **1997**, *104*, 707–713, doi:10.1007/s002570050513.
26. Bianconi, A.; Valletta, A.; Perali, A.; Saini, N.L. Superconductivity of a striped phase at the atomic limit. *Phys. C Supercond.* **1998**, *296*, 269–280, doi:10.1016/s0921-4534(97)01825-x.
27. Bianconi, A. Process of Increasing the Critical Temperature $T_C$ of a Bulk Superconductor by Making Metal Heterostructures at the Atomic Limit. U.S. Patent 6,265,019, 24 July 2001.
28. Bianconi, A. On the possibility of new high $T_C$ superconductors by producing metal heterostructures as in the cuprate perovskites. *Solid State Commun.* **1994**, *89*, 933–936, doi:10.1016/0038-1098(94)90354-9.
29. Bianconi, A.; Missori, M. High $T_C$ superconductivity by quantum confinement. *J. Phys. I Fr.* **1994**, *4*, 361–365, doi:10.1051/jp1:1994100.

30. Cariglia, M.; Vargas-Paredes, A.; Doria, M.; Bianconi, A.; Milošević, M.; Perali, A. Shape-resonant superconductivity in nanofilms: From weak to strong coupling. *J. Supercond. Nov. Magn.* **2016**, *29*, 3081–3086, doi:10.1007/s10948-016-3673-1.
31. Khasanov, R.; Strässle, S.; Di Castro, D.; Masui, T.; Miyasaka, S.; Tajima, S.; Bussmann-Holder, A.; Keller, H. Multiple gap symmetries for the order parameter of cuprate superconductors from penetration depth measurements. *Phys. Rev. Lett.* **2007**, *99*, 237601, doi:10.1103/physrevlett.99.237601.
32. Keller, H.; Bussmann-Holder, A.; Müller, K.A. Jahn–Teller physics and high-$T_C$ superconductivity. *Mater. Today* **2008**, *11*, 38–46, doi:10.1016/s1369-7021(08)70178-0.
33. Bianconi, A. Multiband superconductivity in high $T_C$ cuprates and diborides. *J. Phys. Chem. Solids* **2006**, *67*, 567–570, doi:10.1016/j.jpcs.2005.10.160.
34. Vishik, I.M.; Hashimoto, M.; He, R.H.; Lee, W.S.; Schmitt, F.; Lu, D.; Moore, R.G.; Zhang, C.; Meevasana, W.; Sasagawa, T.; et al. Phase competition in trisected superconducting dome. *Proc. Natl. Acad. Sci. USA* **2012**, *109*, 18332–18337, doi:10.1073/pnas.1209471109.
35. Inosov, D.S.; Park, J.T.; Charnukha, A.; Li, Y.; Boris, A.V.; Keimer, B.; Hinkov, V. Crossover from weak to strong pairing in unconventional superconductors. *Phys. Rev. B* **2011**, *83*, 214520, doi:10.1103/physrevb.83.214520.
36. Bussmann-Holder, A.; Keller, H. Unconventional isotope effects, multi-component superconductivity and polaron formation in high temperature cuprate superconductors. *J. Phys. Conf. Ser.* **2008**, *108*, 012019, doi:10.1088/1742-6596/108/1/012019.
37. Bussmann-Holder, A.; Keller, H.; Khasanov, R.; Simon, A.; Bianconi, A.; Bishop, A.R. Isotope and interband effects in a multi-band model of superconductivity. *New J. Phys.* **2011**, *13*, 093009, doi:10.1088/1367-2630/13/9/093009.
38. Bussmann-Holder, A.; Keller, H. Isotope and multiband effects in layered superconductors. *J. Phys. Condens. Matter* **2012**, *24*, 233201, doi:10.1088/0953-8984/24/23/233201.
39. Perali, A.; Innocenti, D.; Valletta, A.; Bianconi, A. Anomalous isotope effect near a 2.5 Lifshitz transition in a multi-band multi-condensate superconductor made of a superlattice of stripes. *Supercond. Sci. Technol.* **2012**, *25*, 124002.
40. Bianconi, A.; Filippi, M. Feshbach shape resonances in multiband high $T_C$ superconductors. In *Symmetry and Heterogeneity in High Temperature Superconductors*; Bianconi, A., Ed.; Springer: Dordrecht, The Netherlands, 2006; Chapter 2, pp. 21–53.
41. Kristoffel, N.; Rubin, P.; Örd, T. Multiband model of cuprate superconductivity. *Int. J. Mod. Phys. B* **2008**, *22*, 5299–5327, doi:10.1142/s0217979208049443.
42. Deng, S.; Simon, A.; Köhler, J. Flat band–steep band scenario and superconductivity—The case of calcium. *Solid State Sci.* **1998**, *2*, 31–38, doi:10.1016/S1293-2558(00)00105-9.
43. Deng, S.; Simon, A.; Köhler, J. Calcium *d* States: Chemical Bonding of $CaC_6$. *Angew. Chem. Int. Ed.* **2008**, *47*, 6703–6706.
44. Bianconi, A.; Di Castro, D.; Agrestini, S.; Campi, G.; Saini, N.L.; Saccone, A.; De Negri, S.; Giovannini, M. A superconductor made by a metal heterostructure at the atomic limit tuned at the 'shape resonance': $MgB_2$. *J. Phys. Condens. Matter* **2001**, *13*, 7383–7390, doi:10.1088/0953-8984/13/33/318.
45. Agrestini, S.; Di Castro, D.; Sansone, M.; Saini, N.L.; Saccone, A.; De Negri, S.; Giovannini, M.; Colapietro, M.; Bianconi, A. High $T_C$ superconductivity in a critical range of micro-strain and charge density in diborides. *J. Phys. Condens. Matter* **2001**, *13*, 11689–11695, doi:10.1088/0953-8984/13/50/328.
46. Di Castro, D.; Agrestini, S.; Campi, G.; Cassetta, A.; Colapietro, M.; Congeduti, A.; Continenza, A.; Negri, S.D.; Giovannini, M.; Massidda, S.; et al. The amplification of the superconducting $T_C$ by combined effect of tuning of the Fermi level and the tensile micro-strain in $Al_{1-x}Mg_xB_2$. *Europhys. Lett.* **2002**, *58*, 278–284, doi:10.1209/epl/i2002-00634-2.
47. Bianconi, A.; Agrestini, S.; Di Castro, D.; Campi, G.; Zangari, G.; Saini, N.; Saccone, A.; De Negri, S.; Giovannini, M.; Profeta, A.; et al. Scaling of the critical temperature with the Fermi temperature in diborides. *Phys. Rev. B* **2002**, *65*, 174515, doi:10.1103/physrevb.65.174515.
48. Deng, S.; Simon, A.; Köhler, J. Superconductivity in $MgB_2$: A case study of the "flat Band–Steep band" scenario. *J. Supercond.* **2003**, *16*, 477–481, doi:10.1023/A:1023804600927.
49. Deng, S.; Simon, A.; Köhler, J. A "flat/steep band" model for superconductivity. *Int. J. Mod. Phys. B* **2005**, *19*, 29–36, doi:10.1142/s0217979205027895.

50. Bussmann-Holder, A.; Bianconi, A. Raising the diboride superconductor transition temperature using quantum interference effects. *Phys. Rev. B* **2003**, *67*, 132509, doi:10.1103/physrevb.67.132509.
51. Suhl, H.; Matthias, B.T.; Walker, L.R. Bardeen-Cooper-Schrieffer theory of superconductivity in the case of overlapping bands. *Phys. Rev. Lett.* **1959**, *3*, 552–554, doi:10.1103/physrevlett.3.552.
52. Ummarino, G.A.; Gonnelli, R.S.; Massidda, S.; Bianconi, A. Two-band Eliashberg equations and the experimental $T_C$ of the diboride $Mg_{1-x}Al_xB_2$. *Phys. C Supercond.* **2004**, *407*, 121–127, doi:10.1016/j.physc.2004.05.009.
53. Innocenti, D.; Poccia, N.; Ricci, A.; Valletta, A.; Caprara, S.; Perali, A.; Bianconi, A. Resonant and crossover phenomena in a multiband superconductor: Tuning the chemical potential near a band edge. *Phys. Rev. B* **2010**, *82*, 184528, doi:10.1103/physrevb.82.184528.
54. Lifshitz, I.M. Anomalies of electron characteristics of a metal in the high pressure region. *Sov. Phys. JEPT* **1960**, *11*, 1130–1135.
55. Bianconi, A. Feshbach shape resonance in multiband superconductivity in heterostructures. *J. Supercond.* **2005**, *18*, 625–636, doi:10.1007/s10948-005-0047-5.
56. Imada, M.; Misawa, T.; Yamaji, Y. Unconventional quantum criticality emerging as a new common language of transition-metal compounds, heavy-fermion systems, and organic conductors. *J. Phys. Condens. Matter* **2010**, *22*, 164206, doi:10.1088/0953-8984/22/16/164206.
57. Caivano, R.; Fratini, M.; Poccia, N.; Ricci, A.; Puri, A.; Ren, Z.-A.; Dong, X.-L.; Yang, J.; Lu, W.; Zhao, Z.-X.; et al. Feshbach resonance and mesoscopic phase separation near a quantum critical point in multiband FeAs-based superconductors. *Supercond. Sci. Technol.* **2009**, *22*, 014004, doi:10.1088/0953-2048/22/1/014004.
58. Bianconi, A.; Poccia, N.; Ricci, A. Unity in the diversity. *J. Supercond. Nov. Magn.* **2009**, *22*, 526–527, doi:10.1007/s10948-009-0471-z.
59. Innocenti, D.; Valletta, A.; Bianconi, A. Shape resonance at a Lifshitz transition for high temperature superconductivity in multiband superconductors. *J. Supercond. Nov. Magn.* **2011**, *24*, 1137–1143, doi:10.1007/s10948-010-1096-y.
60. Innocenti, D.; Caprara, S.; Poccia, N.; Ricci, A.; Valletta, A.; Bianconi, A. Shape resonance for the anisotropic superconducting gaps near a Lifshitz transition: The effect of electron hopping between layers. *Supercond. Sci. Technol.* **2011**, *24*, 015012, doi:10.1088/0953-2048/24/1/015012.
61. Liu, C.; Palczewski, A.D.; Dhaka, R.S.; Kondo, T.; Fernandes, R.M.; Mun, E.D.; Hodovanets, H.; Thaler, A.N.; Schmalian, J.; Bud'ko, S.L.; et al. Importance of the Fermi-surface topology to the superconducting state of the electron-doped pnictide $Ba(Fe_{1-x}Co_x)_2As_2$. *Phys. Rev. B* **2011**, *84*, 020509, doi:10.1103/physrevb.84.020509.
62. Borisenko, S.V.; Zabolotnyy, V.B.; Kordyuk, A.A.; Evtushinsky, D.V.; Kim, T.K.; Morozov, I.V.; Follath, R.; Büchner, B. One-Sign order parameter in iron based superconductor. *Symmetry* **2012**, *4*, 251–264, doi:10.3390/sym4010251.
63. Kordyuk, A.A.; Zabolotnyy, V.B.; Evtushinsky, D.V.; Yaresko, A.N.; Büchner, B.; Borisenko, S.V. Electronic band structure of Ferro-Pnictide superconductors from ARPES experiment. *J. Supercond. Nov. Magn.* **2013**, *26*, 2837–2841, doi:10.1007/s10948-013-2210-8.
64. Borisenko, S.V.; Evtushinsky, D.V.; Liu, Z.H.; Morozov, I.; Kappenberger, R.; Wurmehl, S.; Büchner, B.; Yaresko, A.N.; Kim, T.K.; Hoesch, M.; et al. Direct observation of spin–orbit coupling in iron-based superconductors. *Nat. Phys.* **2015**, *12*, 311–317, doi:10.1038/nphys3594.
65. Charnukha, A.; Evtushinsky, D.V.; Matt, C.E.; Xu, N.; Shi, M.; Büchner, B.; Zhigadlo, N.D.; Batlogg, B.; Borisenko, S.V. High-temperature superconductivity from fine-tuning of Fermi-surface singularities in iron oxypnictides. *Sci. Rep.* **2015**, *5*, 18273, doi:10.1038/srep18273.
66. Gonnelli, R.S.; Daghero, D.; Tortello, M.; Ummarino, G.A.; Bukowski, Z.; Karpinski, J.; Reuvekamp, P.G.; Kremer, R.K.; Profeta, G.; Suzuki, K.; et al. Fermi-Surface topological phase transition and horizontal Order-Parameter nodes in $CaFe_2As_2$ under pressure. *Sci. Rep.* **2016**, *6*, 26394, doi:10.1038/srep26394.
67. Pustovit, Y.; Kordyuk, A.A. Metamorphoses of electronic structure of FeSe-based superconductors (review article). *Low Temp. Phys.* **2016**, *42*, 995–1007, doi:10.1063/1.4969896.
68. Bianconi, A. Quantum materials: Shape resonances in superstripes. *Nat. Phys.* **2013**, *9*, 536–537, doi:10.1038/nphys2738.

69. Bianconi, A.; Jarlborg, T. Lifshitz transitions and zero point lattice fluctuations in sulfur hydride showing near room temperature superconductivity. *Nov. Supercond. Mater.* **2015**, *1*, 39–47, doi:10.1515/nsm-2015-0006.
70. Jarlborg, T.; Bianconi, A. Breakdown of the Migdal approximation at lifshitz transitions with giant zero-point motion in the $H_3S$ superconductor. *Sci. Rep.* **2016**, *6*, 24816, doi:10.1038/srep24816.
71. Bussmann-Holder, A.; Köhler, J.; Whangbo, M.H.; Bianconi, A.; Simon, A. High temperature superconductivity in sulfur hydride under ultrahigh pressure: A complex superconducting phase beyond conventional BCS. *Nov. Supercond. Mater.* **2016**, *2*, 37, doi:10.1515/nsm-2016-0004.
72. Bussmann-Holder, A.; Köhler, J.; Simon, A.; Whangbo, M.; Bianconi, A. Multigap superconductivity at extremely high temperature: A model for the case of pressurized $H_2S$. *J. Supercond. Nov. Magn.* **2017**, *30*, 151–156, doi:10.1007/s10948-016-3947-7.
73. Perali, A.; Pieri, P.; Strinati, G.C.; Castellani, C. Pseudogap and spectral function from superconducting fluctuations to the bosonic limit. *Phys. Rev. B* **2002**, *66*, 024510.
74. Palestini, F.; Perali, A.; Pieri, P.; Strinati, G.C. Dispersions, weights, and widths of the single-particle spectral function in the normal phase of a Fermi gas. *Phys. Rev. B* **2012**, *85*, 024517.
75. Guidini, A.; Perali, A. Band-edge BCS–BEC crossover in a two-band superconductor: Physical properties and detection parameters. *Supercond. Sci. Technol.* **2014**, *27*, 124002.
76. Shanenko, A.A.; Croitoru, M.D.; Vagov, A.V.; Axt, V.M.; Perali, A.; Peeters, F.M. Atypical BCS-BEC crossover induced by quantum-size effects. *Phys. Rev. A* **2012**, *86*, 033612.
77. Bianconi, A.; Di Castro, D.; Saini, N.L.; Bianconi, G. Superstripes (self organization of quantum wires in high $T_C$ superconductors). In *Phase Transitions and Self-Organization in Electronic and Molecular Networks*; Thorpe, M.F., Phillips, J.C., Eds.; Springer: Boston, MA, USA, 2001; Chapter 24, pp. 375–388.
78. Bianconi, A.; Di Castro, D.; Bianconi, G.; Pifferi, A.; Saini, N.L.; Chou, F.C.; Johnston, D.C.; Colapietro, M. Coexistence of stripes and superconductivity: $T_C$ amplification in a superlattice of superconducting stripes. *Phys. C Supercond.* **2000**, *341–348*, 1719–1722, doi:10.1016/s0921-4534(00)00950-3.
79. Di Castro, D.; Bianconi, G.; Colapietro, M.; Pifferi, A.; Saini, N.L.; Agrestini, S.; Bianconi, A. Evidence for the strain critical point in high $T_C$ superconductors. *Eur. Phys. J. B Condens. Matter Complex Syst.* **2000**, *18*, 617–624, doi:10.1007/s100510070010.
80. Campi, G.; Bianconi, A.; Poccia, N.; Bianconi, G.; Barba, L.; Arrighetti, G.; Innocenti, D.; Karpinski, J.; [1]Zhigadlo, N.D.; Kazakov, S.M.; et al. Inhomogeneity of charge-density-wave order and quenched disorder in a high-$T_C$ superconductor. *Nature* **2015**, *525*, 359–362, doi:10.1038/nature14987.
81. Slezak, J.A.; Lee, J.; Wang, M.; McElroy, K.; Fujita, K.; Andersen, B.M.; Hirschfeld, P.J.; Eisaki, H.; Uchida, S.; Davis, J.C. Imaging the impact on cuprate superconductivity of varying the interatomic distances within individual crystal unit cells. *Proc. Natl. Acad. Sci. USA* **2008**, *105*, 3203–3208, doi:10.1073/pnas.0706795105.
82. Phillips, J.C. Ineluctable complexity of high temperature superconductivity elucidated *J. Supercond. Nov. Magn.* **2014**, *27*, 345–347, doi:10.1007/s10948-013-2308-z.
83. Kresin, V.; Ovchinnikov, Y.; Wolf, S. Inhomogeneous superconductivity and the "pseudogap" state of novel superconductors. *Phys. Rep.* **2006**, *431*, 231–259, doi:10.1016/j.physrep.2006.05.006.
84. Deutscher, G.; de Gennes, P.-G. A spatial interpretation of emerging superconductivity in lightly doped cuprates. *C. R. Phys.* **2007**, *8*, 937–941, doi:10.1016/j.crhy.2007.08.004.
85. Kugel, K.I.; Rakhmanov, A.L.; Sboychakov, A.O.; Poccia, N.; Bianconi, A. Model for phase separation controlled by doping and the internal chemical pressure in different cuprate superconductors. *Phys. Rev. B* **2008**, *78*, 165124, doi:10.1103/physrevb.78.
86. Deutscher, G. The role of Cu-O bond length fluctuations in the high temperature superconductivity mechanism. *J. Appl. Phys.* **2012**, *111*, 112603, doi:10.1063/1.4726157.
87. Bianconi, G. Superconductor-insulator transition in a network of 2d percolation clusters. *Europhys. Lett.* **2013**, *101*, 26003, doi:10.1209/0295-5075/101/26003.
88. Carlson, E.W. Condensed-matter physics: Charge topology in superconductors. *Nature* **2015**, *525*, 329–330, doi:10.1038/525329a.